\documentclass[12pt]{article}
\usepackage{graphicx}

\newcommand\pubnumber{SNSN-XXX-YY}
\newcommand\pubdate{\today}

\def\napoli{
1 - Indiana University, Bloomington, IN 47408, USA
\\ 2 - Universidad Nacional Aut\'{o}noma de M\'{e}xico, M\'{e}xico, D.F. 04510, Mexico
\\ 3 - Thomas Jeﬀerson National Accelerator Facility, VA 23606, USA
\\ 4 - Gettysburg College, Gettysburg, PA 17325, USA
\\ 5- Louisiana State University, Baton Rouge, LA 70803 
\\ 6 - University of Washington/CENPA, Seattle, WA 98195, USA
\\ 7 - North Carolina Central University/TUNL, Durham, NC 27707, USA
\\ 8-University of Dayton, Dayton, OH 45469
\\ 9 - National Institute of Standards and Technology, Gaithersburg, MD 20899, USA
\\ 10 - Hobart William Smith College, Geneva, NY 14456, USA
\\ 11 - Bhabha Atomic Research Center, Trombay, Mumbai 400085, India
\\ 12 - Georgia State University, Atlanta, GA 30303-4106, USA
\\ 13 - Joint Institute for Nuclear Research, 141980 Dubna, Russia
\\ 14 - National High Magnetic Field Laboratory and Florida State University, Dr. Tallahassee, FL 32310-3706, USA
\\ 15- Embry-Riddle Aeronautical University, Daytona Beach, FL 32114

}

\def\Title#1{\begin{center} {\Large #1 } \end{center}}
\def\Author#1{\begin{center}{ \sc #1} \end{center}}
\def\Address#1{\begin{center}{ \it #1} \end{center}}

\newcommand\pubblock{\rightline{\begin{tabular}{l} \pubnumber\\
\pubdate \end{tabular}}}
\newenvironment{Abstract}{\begin{quotation} }{\end{quotation}}
\newenvironment{Presented}{\begin{quotation} \begin{center} 
PRESENTED AT\end{center}\bigskip 
\begin{center}\begin{large}}{\end{large}\end{center} \end{quotation}}
\def\Acknowledgements{\bigskip \bigskip \begin{center} \begin{large}
\bf ACKNOWLEDGEMENTS \end{large}\end{center}}
\begin{document}
\begin{titlepage}
\pubblock

\vfill
\Title{Searches for Exotic Interactions Using Neutron Spin Rotation}
\vfill
\Author{ Chris Haddock$^1$, W. M. Snow$^1$, E. Anderson$^1$, L.~Barr\'{o}n-Palos$^2$, C. D. Bass$^3$, B. E. Crawford$^4$, D. Esposito$^8$, W. Fox$^1$, J. A. Fry$^1$, H. Gardner$^5$, A. T. Holley$^1$, B. R. Heckel$^6$, J. Lieffers$^{15}$, S. Magers$^4$, M. Maldonado-Vel\'{a}zquez$^2$, R. Malone$^4$, D. M. Markoff$^7$, H. P. Mumm$^9$, J. S. Nico$^9$, D. Olek$^10$, Churamani Paudel$^{12}$, S. Penn$^{10}$, 
Prakash Chandra Rout$^{11}$, S. Santra$^{11}$, M. Sarsour$^{12}$, A. Sprow$^4$, H. E. Swanson$^6$, S. Van Sciver$^{14}$, J. Vanderwerp$^1$
}
\Address{\napoli}
\vfill
\begin{Abstract}
Various theories beyond the Standard Model predict new particles with masses in the sub-eV range with very weak couplings to ordinary matter. I present both measured and projected limits on the strengths of two possible interactions that could be mediated by these new particles, and how one may additionally use these results to search for in matter gravitational torsion.
%I present ongoing experimental efforts that use spin-polarized slow neutrons %as a means of investigating possible long range spin-dependent forces.
\end{Abstract}
\vfill
\begin{Presented}
XXXIV Physics in Collision Symposium \\
Bloomington, Indiana, September 16--20, 2014
\end{Presented}
\vfill
\end{titlepage}
\def\thefootnote{\fnsymbol{footnote}}
\setcounter{footnote}{0}

\section{Introduction}

A general classification of interactions between nonrelativistic fermions assuming only rotational invariance uncovered 16 different operator structures involving the spins, momenta, interaction range, and various possible couplings of the particles \cite{Dobrescu}.

The interaction between two fermions mediated by exchange of a spin-1 boson of mass $m_0$ can be generated by a light vector boson $X_{\mu}$ coupling to a fermion $\psi$ with an interaction of the form
$\mathcal{L}_I=\bar{\psi}(g_V\gamma^{\mu}+g_A\gamma^{\mu}\gamma_5)\psi X_{\mu}$ 
where $g_V$ and $g_A$ are the vector and axial couplings. In the nonrelativsitc limit this interaction gives rise to two interaction potentials of interest in our work:

\begin{equation}
V_{PV}=\frac{g_Ag_V}{2\pi}\frac{e^{-r/\lambda}}{r}\vec{\sigma}\cdot\vec{v} 
\end{equation}

\begin{equation}
V_{PC}=\frac{g_A^2}{2\pi}\frac{e^{-r.\lambda}}{r}\vec{\sigma}\cdot(\vec{v}\times\vec{r})
\end{equation}

Here $\lambda=1/m_{X}$ is the interaction range of the new boson of mass $m_X$ and $\vec{\sigma}$ is the spin operator of the polarized particle.\\

Experiments have put upper limits on the strength of the couplings of these possible exotic spin-1 bosons with ordinary matter. Measurements of $V_{PV}$ can also constrain the interaction of a fermion with a background torsion field, which can cause a frame to corkscrew through spacetime (e.g. parallel transport of spin) rather than move along the curve (Einstein GR).

\section{$V_{PV}$ Experimental Arrangement}

To derive the constraint on the strength of the parity violating potential $V_{PV}$ we take advantage of the fact that the potential is proportional to the helicity operator $\vec{\sigma}\cdot\vec{p}$ , and that a particle whose spin is polarized transverse to its velocity may be described a superposition of helicity eigenstates. Thus by sending transverse-polarized neutrons into a target region providing $V_{PV}$, the helicity eigenstates will accumulate different phases and cause the neutron polarization vector to rotate about its momentum.

The rotation angle per unit length $d\phi_{PV}/dL$ of a neutron of wave vector $k_n$ in a medium of density $\rho$ is $d\phi_{PV}/dL=4\pi\rho f_{PV}/k_n$, where $f_{PV}$ is the forward limit of the parity-odd $p$-wave scattering amplitude. Using the Born approximation to express $f_{PV}$ in terms of the parameters of the potential $V_{PV}$, the following relation is obtained:

\begin{equation}
\frac{d\phi_{PV}}{dL}=4g_Vg_A\rho\lambda^2
\end{equation}

A beam of low-energy neutrons passes through a liquid helium target located between a crossed polarizer/analyzer pair. Neutrons initially travel through a vertically aligned supermirror polarizer and are guided into a field-free region where they pass through target chambers filled with liquid $^4He$. While traveling through the chamber the vertically polarized neutrons can rotate about their momentum due to the $V_{PV}$ interaction with the nucleons in $^4He$.

Target chambers located in front and back of a $\pi$-coil are alternately filled with liquid helium. This $\pi$-coil reverses the horizontal transverse component of neutron polarization by Larmour precession about the vertical axis. In combination with modulating the $^4He$ between the chambers, the PV signal is doubled while external magnetic field rotations are cancelled to first order.\\ 

\begin{figure}[htb]
\centering
\includegraphics[scale=.32]{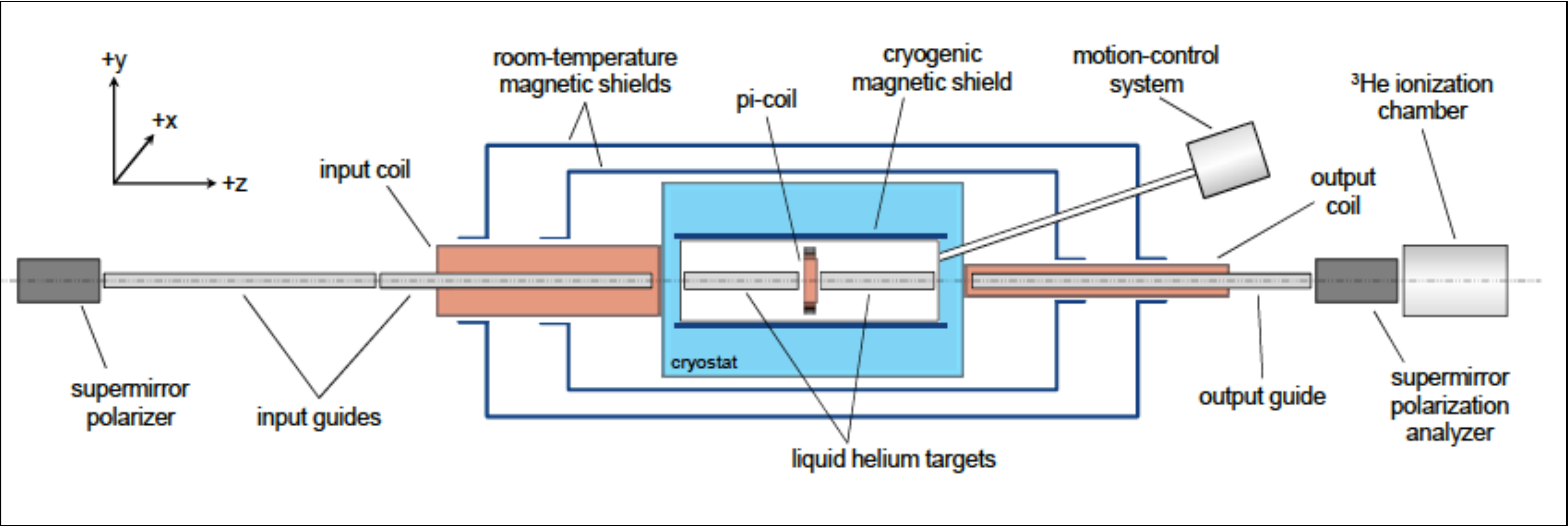}
\caption{Experimental apparatus used to measure $V_{PV}$. Neutrons enter from the left.}
\label{fig:magnet}
\end{figure}

The measured upper bound on the parity-odd spin rotation angle per unit length in liquid $^4He$ at a temperature of 4 K from this experiment is\\

\begin{equation}
\frac{d\phi_{PV}}{dL}=+1.7\pm 9.1(stat)\pm 1.4(syst)\times 10^{-7}rad/m
\end{equation}

\vspace{1cm}

Combining the result in (4) with the epxression given in (3) generates the following exclusion plot:

\pagebreak

\begin{figure}[htb]
\centering
\includegraphics[scale=.45]{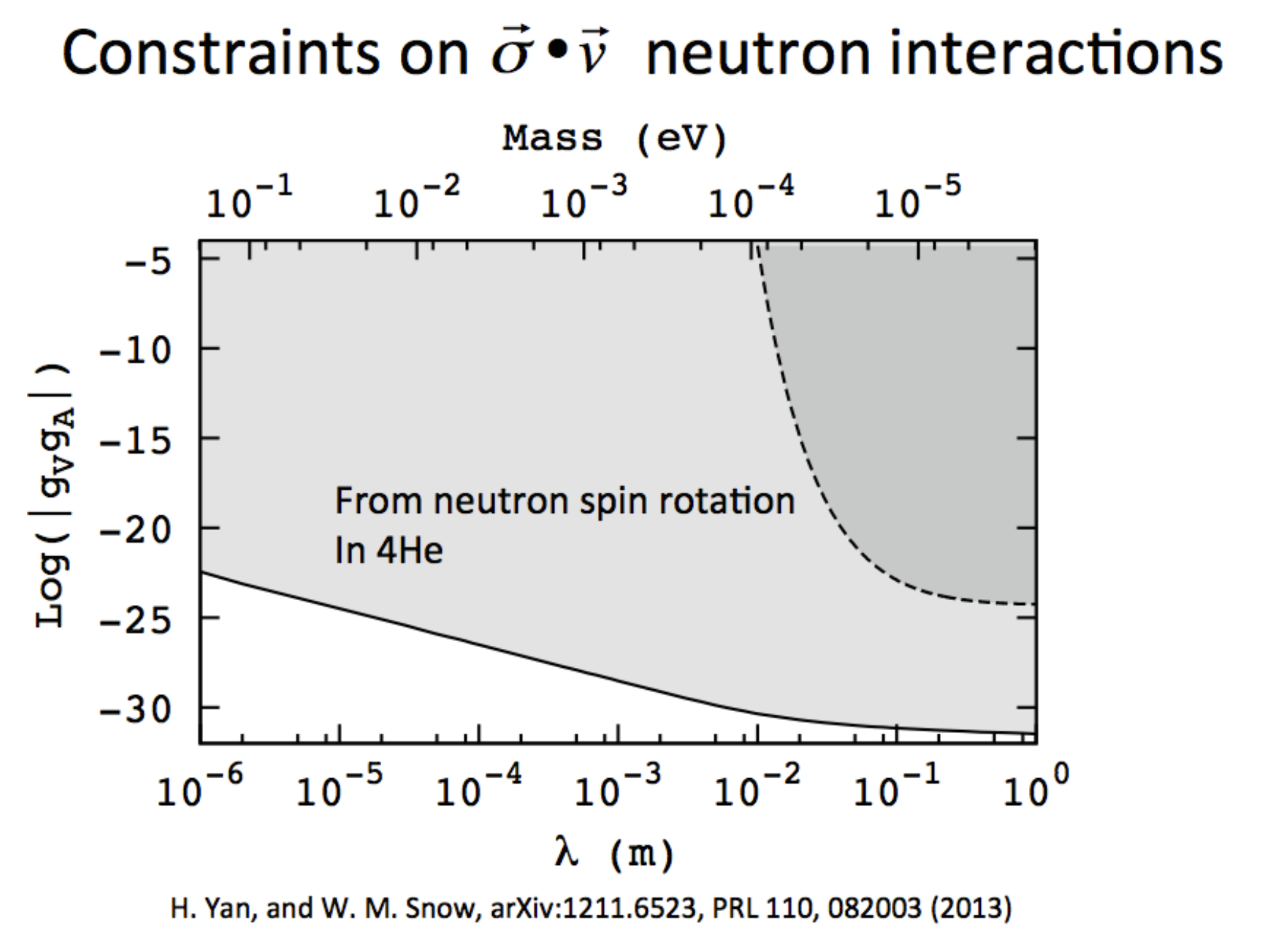}
\caption{Upper limits placed on parity-odd neutron interactions from neutron spin rotation in $^4He$.}
\label{fig:magnet}
\end{figure}

\section{In-Matter Torsion}

A parity-odd spin-dependent interaction may also indicate the presence of gravitational torsion~\cite{torsion}. This phenomenon can be described by an external torsion field 
$\tau^{\alpha}_{\mu\nu}$ generated by the spin density of the liquid $^4He$ affecting the spin state of the polarized neutron beam. The general structure of the nonrelativistic neutron Hamiltonian can be written as:

\begin{equation}
\mathcal{H}=\frac{\vec{P}^2}{2m}+\delta\vec{b}\cdot\vec{\sigma}
\end{equation}

where $\delta\vec{B}$ is determined by the background torsion field. In our present example with $^4He$ the torsion effects produced are isotropic and on a macroscopic scale. The background torsion then takes the form

\begin{equation}
\delta\vec{b}=\left(\frac{\zeta}{m}\right)\vec{p}
\end{equation}

where $\zeta$ is a model-dependent parameter that depends on the rotationally invariant pieces of the torsion tensor $\tau^{\alpha}_{\mu\nu}$. A non-zero $\zeta$ would indicate an interaction between a torsion field arising from the spin density of the fermions in $^4He$ and the beam of spin-polarized neutrons.

As the form of this potential is the same as that of $V_{PV}$, we may reinterpret the results of our $V_{PV}$ measuremeant by deriving the analagous relation between the spin rotation angle per unit length for polarized neutrons passing though $^4He$ and the torsion parameter $\zeta$ :

\begin{equation}
\frac{d\phi}{dL}=2\zeta
\end{equation}\vspace{.1cm}

Applying the result from (4) gives:

\begin{equation}
\zeta=+1.7\pm 9.1(stat)\pm 1.4(syst)\times 10^{-23}GeV
\end{equation}\vspace{.1cm}

 Our constraint is the first experimental search to our knowledge for in-matter gravitational torsion.

\section{$V_{PC}$ Experimental Arrangement }

By making slight modifications to the neutron polarimeter designed to measure parity-odd spin rotations, we can also measure the parity-even rotations that neutrons would experience from $V_{PC}$ . Consider a beam of polarized neutrons moving parallel to a flat plate of homogeneous material in the presence of an interaction $V_{PC}\propto \vec{\sigma}\cdot(\vec{v}\times\vec{r})$, where $\sigma$ is the spin of the polarized neutrons, $\vec{v}$ is their velocity, and $\vec{r}$ is their position from the plate. By viewing the $(\vec{v}\times\vec{r})$ term as a pseudo-magnetic field directed transverse to the neutron velocity, it is apparent that the neutron polarization will precess in the longitudinal plane as it passes over the plate:

\begin{figure}[htb]
\centering
\frame{\includegraphics[scale=.3]{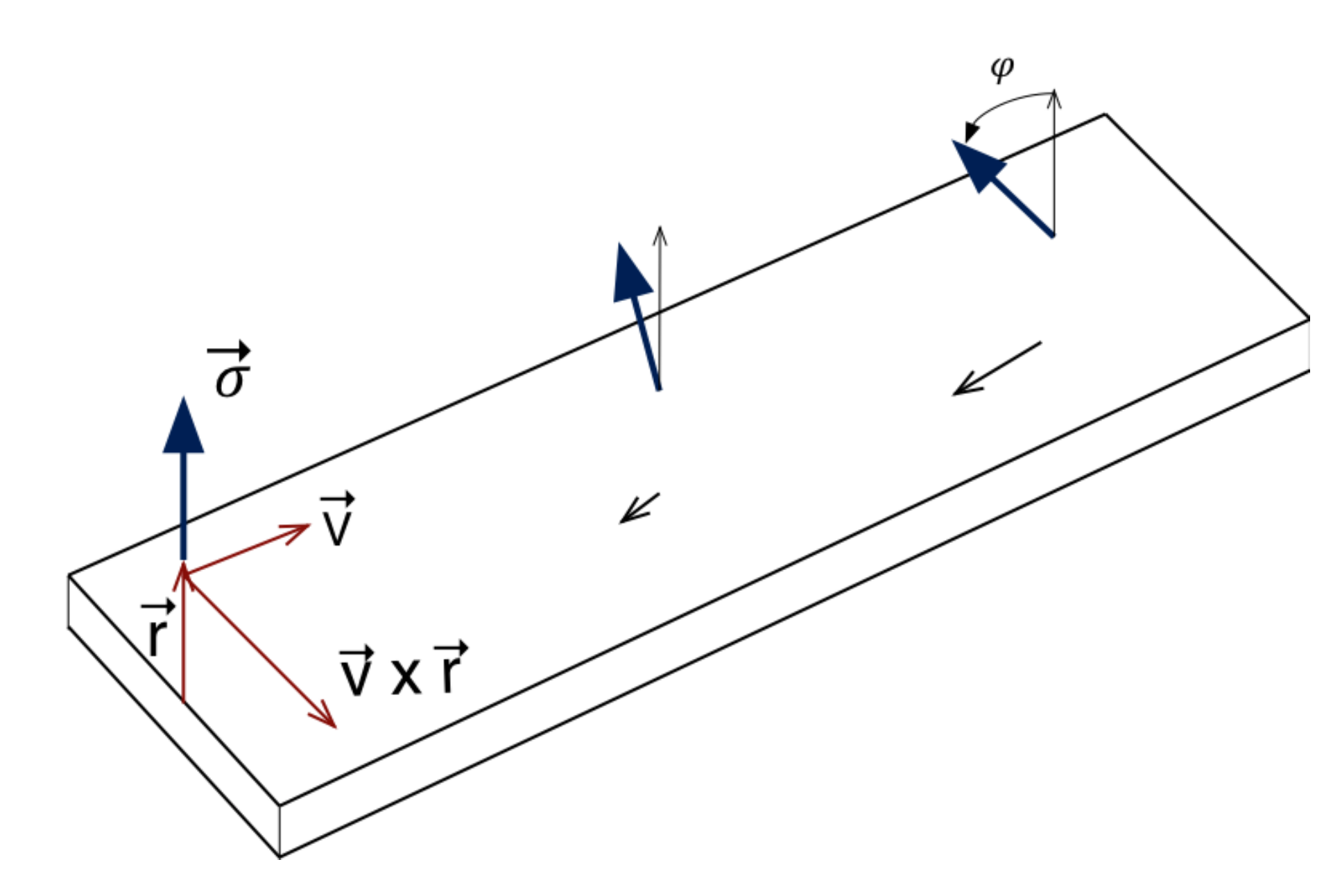}}
\caption{Neutron polarization rotates about an axis transverse to its velocity in the case of $V_{PC}$.}
\label{fig:magnet}
\end{figure}

By assembling an array of plates of different nucleon density so that neutrons traveling in an empty region will always see a nucleon density gradient, the neutrons will undergo a net precession in the presence of this interaction.  This can limit  the coupling constant product $g_A^{2}$. Monte Carlo simulations have been performed and suggest that a neutron experiment which improves present limits by many orders of magnitude for interactions ranges below 1mm is feasible.

\begin{figure}[htb]
\centering
\frame{\includegraphics[scale=.5]{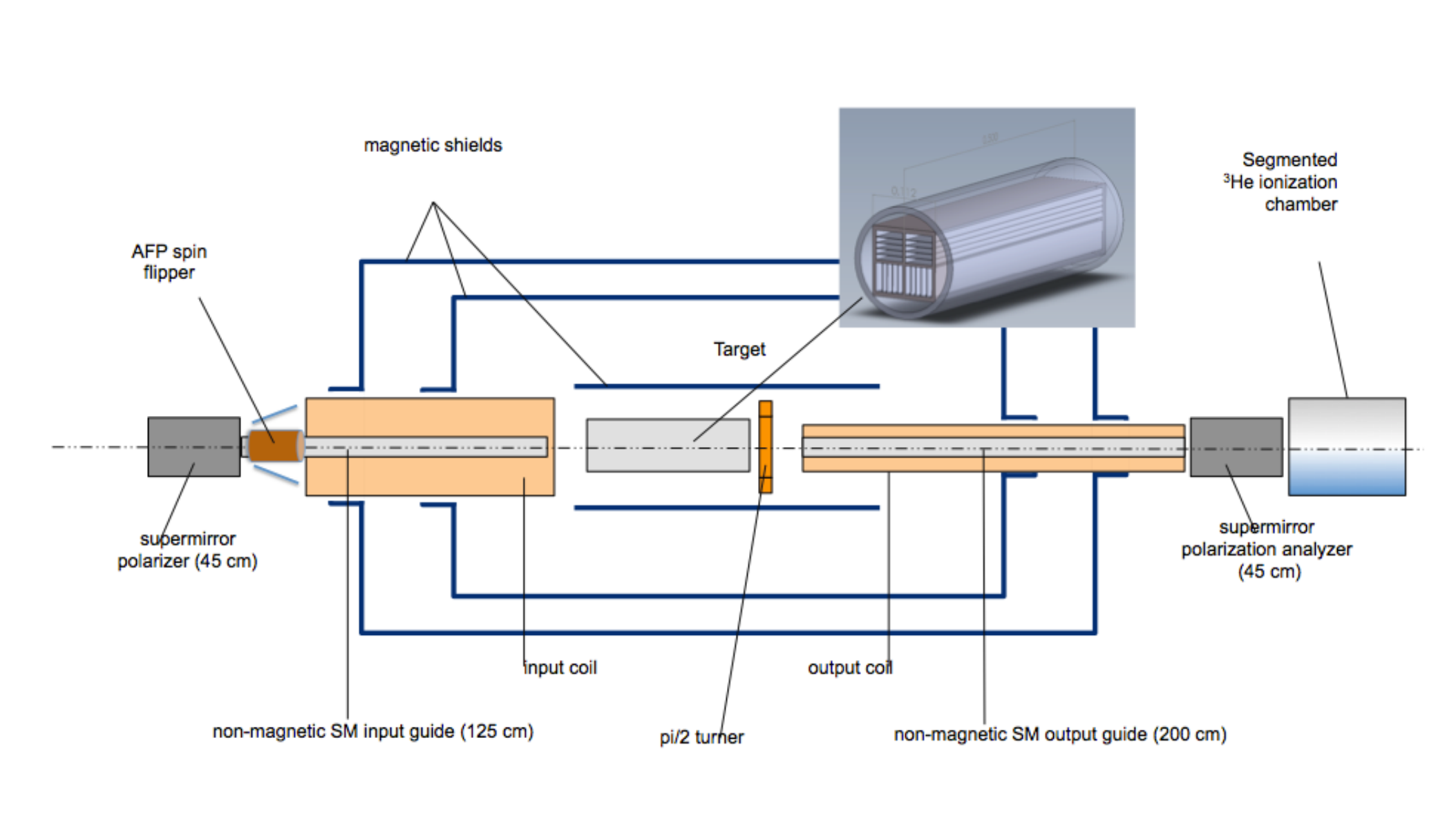}}
\caption{Experimental apparatus to measure $V_{PC}$ (modified $V_{PV}$ apparatus)}
\label{fig:magnet}
\end{figure}

\Acknowledgements
This work was supported in part by NSF PHY-0457219, NSF PHY-0758018, DE-AI02-93ER40784, DE-SC0010443, and DE-FG02-95ER40901. We acknowledge the support
of the National Institute of Standards and Technology, US Department of Commerce, in providing the neutron facilities used in this work. H. Gardiner, D. Esposito, and J. Lieffers acknowledge support from the NSF Research Experiences for Undergraduates program NSF PHY-1156540. W. M. Snow, E. Anderson, J. Fry, C. Haddock, and A. T. Holley acknowledge support from the Indiana University Center for Spacetime Symmetries.

\end{document}